\documentclass[11pt]{article}

\usepackage{amssymb,amsthm,amscd,array}

\usepackage[cyr]{aeguill}     

\let\ssection=\section
\renewcommand{\section}{\setcounter{equation}{0}\ssection}

\newcommand{\bbR}{\mathbb{R}}

\newcommand{\Diff}{\mathrm{Diff}}

\newcommand{\rg}{\mathrm{g}}

\newcommand{\half}{\frac{1}{2}}

\newtheorem{thm}{Theorem}[section]

\title{Schwarzian derivative\\ and\\ Numata Finsler structures}
\author{
C. DUVAL\footnote{mailto: duval@cpt.univ-mrs.fr}\\\\
Centre de Physique Th\'eorique, CNRS, 
Luminy, Case 907\\ 
F-13288 Marseille Cedex 9 (France)\footnote{ 
UMR 6207 du CNRS associ\'ee aux 
Universit\'es d'Aix-Marseille I et II et Universit\'e du Sud Toulon-Var; Laboratoire 
affili\'e \`a la FRUMAM-FR2291
}
}
\date{}    

\begin{document}

\maketitle

\abstract{The flag curvature of the Numata Finsler structures is shown to admit a non\-trivial prolongation to the one-dimensional case, revealing an unexpected link with the Schwarzian derivative of the diffeomorphisms associated with these Finsler structures.}

\bigskip

Mathematics Subject Classification 2000: 58B20, 53A55

\baselineskip=18pt

\section{Finsler structures in a nutshell}

\subsection{Finsler metrics}

A Finsler structure is a pair $(M,F)$ where $M$ is a smooth, $n$-dimensional, manifold and $F:TM\to\bbR^+$ a given function whose restriction to the slit tangent bundle $TM\!\setminus\!M=\{(x,y)\in{}TM\,\vert\,y\in{}T_xM\!\setminus\!\{0\}\}$ is strictly positive, smooth, and positively homogeneous of degree one, i.e., $F(x,\lambda y)=\lambda F(x,y)$ for all $\lambda>0$; one furthermore demands that the $n\times{}n$ vertical Hessian matrix with entries 
$\rg_{ij}(x,y)=\left(\half{}F^2\right)_{y^i y^j}$
be positive definite, $(\rg_{ij})>0$. See \cite{BCS}. These quantities are (positively) homogeneous of degree zero, and the fundamental tensor
\begin{equation}
\rg=\rg_{ij}(x,y)dx^i\otimes{}dx^j
\label{g}
\end{equation}
defines a \textit{sphere's worth of Riemannian metrics} on each $T_xM$ parametrized by the direction of $y$. See \cite{BR}.


The distinguished ``vector field'' 
\begin{equation}
\ell=\ell^i\frac{\partial}{\partial{x^i}},
\qquad
\mathrm{where}
\qquad
\ell^i(x,y)=\frac{y^i}{F(x,y)},
\label{ell}
\end{equation}
actually a section of $\pi^*(TM)$ where $\pi:TM\!\setminus\!{}M\to{}M$ is the natural projection, is such that 
$\rg(\ell,\ell)
=1$.

\goodbreak

There is a wealth of Finsler structures, apart from the special case of Riemannian structures $(M,\rg)$ for which $F(x,y)=\sqrt{\rg_{ij}(x)y^iy^j}$. For instance, the so-called Randers metrics
\begin{equation}
F(x,y)=\sqrt{a_{ij}(x)y^iy^j}+b_i(x)y^i
\label{Randers}
\end{equation}
satisfy all previous requirements if $a=a_{ij}(x)dx^i\otimes dx^j$ is a Riemann metric and if the $1$-form $b=b_i(x)dx^i$ is such that $a^{ij}(x)b_i(x)b_j(x)<1$ for all $x\in{}M$.

\goodbreak

\subsection{Flag curvature}

Unlike the Riemannian case, there is no canonical linear Finsler connection on $\pi^*(TM)$. An example, though, is provided by the Chern connection $\omega^i_j=\Gamma^i_{jk}(x,y)dx^k$ which is uniquely defined by the following requirements \cite{BCS}: (i) it is symmetric, $\Gamma^i_{jk}=\Gamma^i_{kj}$, and (ii) it \textit{almost} transports the metric tensor, i.e., $dg_{ij}-\omega^k_ig_{jk}-\omega^k_jg_{ik}=2C_{ijk}\delta{y^k}$, with $\delta{y^i}=dy^i+N^i_jdx^j$, where the $N^i_j(x,y)=\Gamma^i_{jk}y^k$ are the components of the non linear connection associated with the Chern connection, and the
$
C_{ijk}(x,y) 
=\half\left(\rg_{ij}\right)_{y^k}
$
those of the Cartan tensor, specific to Finsler geometry.

Using the ``horizontal covariant derivatives'' $\delta/\delta{x^i}=\partial/\partial{x^i}-N^j_i\partial/\partial{y^j}$, one expres\-ses the (horizontal-horizontal part of the) Chern curvature by
\begin{equation}
R^{\ i}_{j\ kl}
=
\frac{\delta}{\delta{x^k}}\Gamma^i_{jl}+\Gamma^i_{mk}\Gamma^m_{jl}-(k\leftrightarrow{}l),
\label{R}
\end{equation}
and the \textit{flag curvature} (associated with the flag $\ell\wedge{}v$ defined by $v\in{}T_xM$) by
\begin{equation}
K(x,y,v)
=
\displaystyle
\frac{R_{ik}v^iv^k}{\rg(v,v)-\rg(\ell,v)^2},
\qquad
\mathrm{where}
\qquad
R_{ik}=\ell^jR_{jik\ell}\,\ell^\ell.
\label{K}
\end{equation}

One says that a Finsler structure is \textit{of scalar curvature} if $K(x,y,v)$ does not depend on the vector $v$, i.e., if
\begin{equation}
R_{ik}
=
K(x,y)h_{ik},
\label{scalarK}
\end{equation}
with $h_{ik}=\rg_{ik}-\ell_i\ell_k$ the components of the ``angular metric'', where $\ell_i=\rg_{ij}\ell^j(=F_{y^i})$. See \cite{BCS,BR}.

\section{Numata Finsler structures}

\subsection{The Numata metric}

Numata \cite{Num} has proved that metrics of the form $F(x,y)=\sqrt{q_{ij}(y)y^iy^j}+b_i(x)y^i$, on $TM$ where $M\subset{}\bbR^n$, with $(q_{ij})>0$ and $db=0$ are, indeed, of scalar curvature. See~\cite{BR}.

\goodbreak

Of some interest is the special case $q_{ij}=\delta_{ij}$ and $b=df$ with $f\in{}C^\infty(M)$, viz.,
\begin{equation}
F(x,y)=\sqrt{\delta_{ij}y^iy^j}+f_{x^i}y^i,
\label{Numata}
\end{equation}
where 
\begin{equation}
M=\Big\{x\in\bbR^n\,\Big\vert\,\sum_{i=1}^n{f^2_{x^i}}<1\Big\}.
\label{M}
\end{equation}

\goodbreak

The computation of the flag curvature of this particular Randers metric (\ref{Randers}) can be found in \cite{BCS} and yields
\begin{equation}
K(x,y)
=
\frac{3}{4}\frac{1}{F^4}\left(f_{x^ix^j}y^iy^j\right)^2
-
\frac{1}{2}\frac{1}{F^3}\,f_{x^ix^jx^k}y^iy^jy^k.
\label{KNumata}
\end{equation}

\subsection{Flag curvature \& Schwarzian derivative}

The expression (\ref{KNumata}) of the flag curvature of the Numata metric (\ref{Numata}) holds for $n\geq2$. 

If $n=1$, the left-hand side of~(\ref{scalarK}) vanishes along with the curvature (\ref{R}), while its right-hand vanishes as well since the angular metric has rank zero. For this particular dimension, Equation (\ref{scalarK}) trivially holds true, but tells, however, nothing about the flag curvature $K(x,y)$.

At this stage, it is worth noting that (\ref{KNumata}) indeed admits a prolongation to the one-dimensional case; it is therefore tempting to specialize its expression for $n=1$.

Suppose, thus, that $M\subset{}S^1$ is a nonempty open subset (\ref{M}), so that  we have $TM\!\setminus\!M=T_+M\bigsqcup{}T_-M$, where $T_\pm{}M=M\times\bbR^\pm_*$. The metric (\ref{Numata}) then reads
\begin{equation}
F(x,y)=\vert{}y\vert+f'(x)y,
\label{n=1Numata}
\end{equation}
using an affine coordinate, $x$, on $S^1$, with $-1<f'(x)<+1$ (see (\ref{M})); its restrictions to $T_\pm{}M$ are given by $F_\pm(x,y)=\varphi'_\pm(x)y>0$, 
where
\begin{equation}
\varphi'_\pm(x)=f'(x)\pm1,
\label{phipm}
\end{equation}
implying
$\varphi_\pm\in\Diff_\pm(S^1)$,
with $\vert\varphi'_\pm(x)\vert<2$ (all $x\in{}M$).

The Numata metric~(\ref{n=1Numata}) on $T_+M$, say, is thus associated, via (\ref{phipm}), to orientation-preserving diffeomorphisms~$\varphi$ of $S^1$ such that $0<\varphi'(x)<2$ (all $x\in{}M$).
Given such a $\varphi\in\Diff_+(S^1)$, the fundamental tensor (\ref{g}) retains the form $\rg=\varphi'(x)^2dx^2$ and is, naturally, a Riemannian metric on $M$.

Rewriting Equation (\ref{KNumata}) for $T_+M$, and bearing in mind that
$y=F(x,y)/\varphi'(x)$, we readily find that $K(x,y)$ is actually independent of $y$, namely
\begin{equation}
K(x)=-\half\frac{1}{\varphi'(x)^2}\,S(\varphi)(x),
\label{SchwarzNumata}
\end{equation}
where
\begin{equation}
S(\varphi)(x)=\frac{\varphi'''(x)}{\varphi'(x)}-\frac{3}{2}\left(\frac{\varphi''(x)}{\varphi'(x)}\right)^2
\label{S}
\end{equation}
denotes the \textit{Schwarzian derivative} \cite{OT} of the  diffeomorphism $\varphi$ of $S^1$. The argument clearly still holds, \textit{mutatis mutandis}, for orientation-reversing diffeomorphisms of $S^1$.

\goodbreak

We have thus proved the
\begin{thm}
The Numata Finsler structure $(M,F)$, with metric~$F$ given by (\ref{n=1Numata}) where $M\subset{}S^1$ is defined by (\ref{M}), induces a Riemannian metric, $\mathsf{g}(\varphi)=\varphi^*(dx^2)$, where $\varphi\in\Diff(S^1)$ is as in~(\ref{phipm}). The flag curvature (\ref{KNumata}) admits a prolongation to this one-dimensional case and retains the form
\begin{equation}
K=-\half\frac{\mathsf{S}(\varphi)}{\mathsf{g}(\varphi)},
\label{KS/g}
\end{equation}
where $\mathsf{S}(\varphi)=S(\varphi)(x)dx^2$ is the Schwarzian quadratic differential of $\varphi\in\Diff(S^1)$.
\end{thm}

As an illustration, the one-dimensional Numata Finsler structures of constant flag curvature are associated, through (\ref{phipm}), to the solutions $\varphi$ of (\ref{KS/g}) for $K\in\bbR$, viz., $\varphi_\pm(x)=K^{-\half}\arctan(K^{\half}(a x + b)/(c x + d))$ where $a,b,c,d\in\bbR$ with $ad-bc=\pm1$.

Let us mention another instance where the Schwarzian derivative is associated with curvature, namely the geometry of curves in Lorentzian surfaces of constant curvature~\cite{DO}.

\medskip
Discussions with P. Foulon are warmly acknowledged.



\end{document}